\documentstyle[prl,aps,epsf,twocolumn]{revtex}
\begin{document}
\draft
\twocolumn[\hsize\textwidth\columnwidth\hsize\csname @twocolumnfalse\endcsname
\preprint{Submitted to Phys. Rev. B}
\title{The Structure of Fractional Edge States: A Composite Fermion Approach}
\author {Dmitri B. Chklovskii}
\address{
Physics Department, Massachusetts Institute of Technology, Cambridge, MA 02139
\\
 Present address: Physics Department, Harvard University, Cambridge, MA 02138}
\maketitle

\begin{abstract}

I study the structure of the two-dimensional electron gas edge in the quantum
Hall regime using the composite
fermion  approach. The electron density distribution
and the composite fermion energy  spectrum are obtained numerically in Hartree
approximation for bulk filling factors $\nu=1,1/3,2/3,1/5$. For a very sharp
edge of the $\nu=1$ state the one-electron picture is valid.
 As the edge width $a$ is increased the
density distribution shows features related to the fractional states and new
fractional
 channels appear in pairs. For a very smooth edge I find quasiclassically
 the number of channels $p\sim\sqrt{a/l_H}$, where $l_H$ is the magnetic
length.

\end{abstract}

]

\section{Introduction}
The concept of edge states was introduced originally in the framework
of the integer quantum Hall effect (IQHE).\cite{Halperin82} It is based on
 the fact that the two-dimensional electron gas (2DEG) at an
integer Landau level filling factor $\nu$ can support gapless excitations
 only at the edge. Hence the transport properties can be understood without
explicitly
considering the bulk of the sample. The edge of the 2DEG is created by the
confinement
 potential which bends the Landau levels up. The intersection of each
Landau level with the Fermi energy gives origin to a chiral one-dimensional
 edge channel. Naturally, the total number of channels within the one-electron
theory is given by the bulk filling factor.
Experiments with non-ideal contacts\cite{vanWees,Komiyama,Alphenaar} have
confirmed the
 existence of separate edge channels corresponding to different Landau levels.
The quantized multi-probe resistance was found to be in agreement with the
 predictions of the Landauer-B\"uttiker formalism\cite{Buttiker}
 based on the one-electron picture. However, a number of more sophisticated
 experiments were impossible to explain on the one-electron footing.
\cite{Alphenaarpre,Kouwenhoven2/3}

Chklovskii {\it et al.}\cite{Chklovskii1} developing earlier
 works\cite{Efros,Luryi,Kane,Beenakker,Chang,McEuen} have shown that in a
 realistic confinement potential edge channels (compressible strips)
 have finite width as a result  of electron-electron interactions.
The edge channels being not strictly one-dimensional brings up the problem of
 their  internal structure. In the two limits of very sharp and very smooth
confinement this problem can be tackled. When the confinement potential
 is very sharp it dominates over the electron-electron interactions and the
 single-electron theory must be valid. However it is unclear how to implement
this situation experimentally. In a very smooth confinement potential
 a compressible strip breaks up into a number of fractional states with
fractional
edge channels between them.\cite{Beenakker,Chang} This explains the observation
of
 the fractionally quantized Hall conductance for integer bulk filling factor in
 devices with non-ideal contacts.\cite{Kouwenhoven2/3}

The purpose of this paper is to study the evolution of the edge channel
structure
 as the  smoothness of the confinement is changed continuously. The complexity
of
 the problem arises from the strongly correlated nature of the electron state.
 Any theory that resolves this problem should be able to describe the
one-electron
limit of very sharp confinement as well as the formation of the incompressible
fractional states in a very smooth confinement. In this paper I present such  a
 theory which also predicts the number of fractional edge channels as the
function
of the confinement sharpness. The theory is based on the composite fermion
 approach.\cite{Jain,Lopez,Kalmeyer,HLR} The advantage of this approach
 is that the important electron-electron correlations are automatically built
into
 the single-particle (Hartree) approximation for composite fermions.

\begin{figure}
\epsfxsize=3truein
\hskip 0.0truein \epsffile{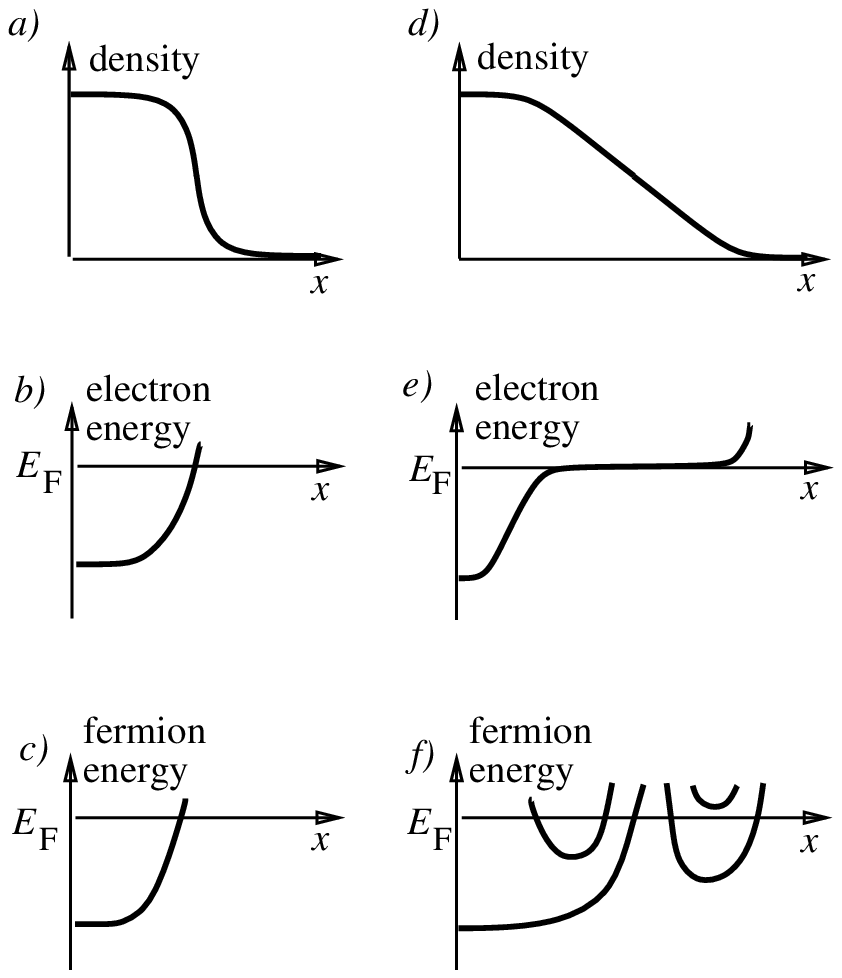}

 { FIG 1: Schematic density distribution and energy spectra for the sharp and
smooth
edge of the $\nu=1$
 state. Energy is plotted as a function of the guiding center position of
momentum eigenstates.
a) Density distribution for the sharp edge.
b) Electron energy spectrum for the sharp edge.
c) Composite fermion energy spectrum for the sharp edge.
d)  Density distribution for the smooth edge.
e) Electron energy spectrum for the smooth edge.
f) Composite fermion energy spectrum for the smooth edge.}
\label{fig:schematic}
\end{figure}

The main idea can be demonstrated by considering the evolution of the $\nu=1$
edge as the confinement potential varies from the very sharp to the very smooth
limit.(see Fig.\ref{fig:schematic}) The confinement potential can be
characterized
by a single parameter $a$
 which represents the width of the compressible liquid strip where the filling
 factor drops from one to zero.

When the confinement potential is very sharp ($a\sim l_H$) the one-electron
 theory must be valid, implying that there is just a single
one-dimensional channel, Fig.\ref{fig:schematic}b. This result can be also
understood in terms of composite
 fermions. Attachment of two flux quanta maps the $\nu=1$ electron state on
 a single filled Landau level for composite fermions with the magnetic
field direction reversed. At the edge this level bends up and intersects the
 Fermi energy forming a single channel, Fig.\ref{fig:schematic}c. Higher Landau
levels for composite fermions remain empty.

As the confinement gets smoother (larger $a$) a finite width region with
the partially filled  Landau level emerges, Fig.\ \ref{fig:schematic}d.
One-electron theory
 is useless in this case since it predicts a huge degeneracy,
Fig.\ref{fig:schematic}e.
 The composite  fermion approach, on the other hand, yields an interesting
picture,
 Fig.\ref{fig:schematic}f.\cite{Chklovskii3}
 The effective magnetic field seen by composite fermions is proportional to
the deviation of the filling factor from $1/2$. Hence the compressible strip
 with the  filling factor dropping  from one to zero results in the effective
 magnetic field varying between minus and plus the external magnetic field
value.
 The problem is reduced then to a particle in a varying effective
magnetic field.\cite{Muller,Chklovskii3,Lee} Composite
fermion energy bands descend in the regions of weak effective magnetic field
because
of the reduction in the effective cyclotron energy. As parameter $a$ is
increased
the gradient of the effective magnetic field gets smaller and higher
energy bands intersect the Fermi energy, Fig.\ref{fig:schematic}f. When each
consecutive
band touches the Fermi
 energy a pair of fractional channels is born. They are composed out of states
 moving in the opposite directions. When the channels are just
born it is impossible to identify them with any particular fraction.
 But for sufficiently large $a$, the channels become separated by the
fractional
filling factor regions.\cite{Beenakker,Chang}

In this paper I show that the above picture is self-consistent within Hartree
 approximation for composite fermions. The electron density distribution and
the
 composite fermion energy spectrum are found for several bulk filling factors.
 The model used in the numerical solution
 and the formalism of Hartree approximation are explained in
Sec.\ref{sec:model}.
  In Sec.\ref{sec:res1/3} results of the
 Hartree approximation for simple fractions are compared against the existing
Laughlin wavefunction calculations to check the reliability of the method.
Encouraged
 by the ability of the Hartree calculation to reproduce the essential features
of
the density distribution I apply the method
to study the evolution of the $\nu=1$ edge in Sec.\ref{sec:res1}. In
Sec.\ref{sec:res2/3}
 the composite fermion approach is used to address the longstanding
 problem of the $\nu=2/3$ edge.  In short, I find that
MacDonald\cite{MacDonald}
and Chang-Beenakker\cite{Chang,Beenakker} models describe sharp
 and smooth confinement, correspondingly, in agreement with the conclusions of
 Meir\cite{Meir} and Brey\cite{Brey}.

In the limit of a very smooth confinement the number of occupied
composite fermion bands is large, calling for the quasiclassical
approximation. This approach is used in Sec.\  \ref{sec:classical}
to find the number of fractional channels in the large $a$ limit.
It also gives the highest fractional Hall state in the principal
sequence $p/(2p+1)$\cite{Jain}
that survives a density gradient determined by the lengthscale $a$.
The dependence of the highest $p$ on the Landau level number is found.

\section{Model and method.}
\label{sec:model}
\begin{figure}
\epsfxsize=3truein
\hskip 0.0truein \epsffile{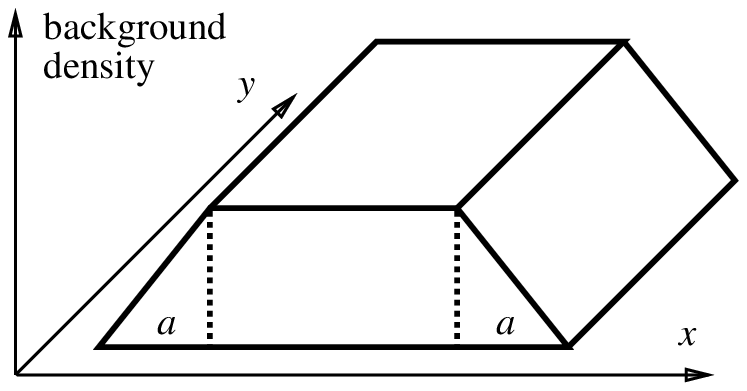}

{ FIG 2: Neutralizing background charge density used to model the realistic
confinement. The sharpness
of the edge is determined by parameter $a$ which is defined by the zero
magnetic field electron
density distribution. The distance between the two edges is chosen large enough
to eliminate interference between them.}
\label{fig:model}
\end{figure}

In GaAs heterostructures the 2DEG edge is most commonly defined either by
 a negatively biased gate  on top of the device or by the chemical etching
 process.\cite{BvH} In both confinement schemes the electron density
 distribution at the edge has to be  determined self-consistently.
\cite{Chklovskii1,Gelfand} For the purpose of studying the structure of
the edge channels I consider a simplified model instead of the actual
confinement scheme.
 The  2DEG is placed on a positive background, the density
of which mimics the electron density in the absence of a magnetic field.
 This model is supported by the smallness of the magnetic
 field-caused electron density redistribution.\cite{Chklovskii1}
For computational reasons I consider a quantum wire with two edges.
Yet another simplification has to be made in order to keep the bulk filling
factor constant while the confinement potential is varied. Instead of
a realistic density profile\cite{Larkin} I consider the positive background
 of the trapezoidal form in the $x-z$ plane
and translationally invariant in the $y$ direction.(Fig.\ref{fig:model})
 The density in the center corresponds to various bulk filling factors.
 The separation between the two edges in the $x$ direction is chosen large
enough to keep  the  interference between them negligible.
In order to apply results of the
calculation to a particular experimental situation one has to estimate the
typical
 distance $a$ on which the electron density drops from its bulk value to zero.
 The trapezoidal background model with the width
$a$ of the gradient regions is a good approximation for most  purposes.

The above model has been used to study integer edge channels in Hartree
approximation\cite{BreyHF}, and the edge reconstruction of the
$\nu=1$ state.\cite{Chamon} Both works did not treat electron-electron
correlations which are explicitly present in this calculation.
The model used by Brey\cite{Brey} is essentially the same except that
he places an infinite potential wall at the edge of the trapezoid. I believe
that such a boundary condition imposes an unrealistically strong confinement.
In my model the infinite potential wall is positioned far enough from the edge
of the
trapezoid to make the solution insensitive to its exact location.

The advantage of using the composite fermion approach is in transforming
a system of strongly correlated electrons to the system of weakly
interacting quasiparticles. Formally this can be realized in
the Hartree approximation for {\it composite fermions}.
The starting point is the electron Hamiltonian,
\begin{equation}
H=\sum_{i=1}^N \frac{({\bf p}_i+e/c{\bf A}_i)^2}{2m_B} + \sum_{i\neq j}
 V({\bf r}_i-{\bf r}_j) +
 \sum_{i=1}^N U({\bf r}_i),
\end{equation}
where ${\bf A}$ is the external vector-potential, $m_B$ is the electron band
mass,
$V$ is the electron-electron interaction, and $U$ is the external potential.
After the singular gauge transformation the Hamiltonian has the form
\begin{equation}
H=\sum_{i=1}^N \frac{({\bf p}_i+e/c{\bf A}_i-{\bf a}_i)^2}{2m_B} +
 \sum_{i\neq j} V({\bf r}_i-{\bf r}_j) +
 \sum_{i=1}^N U({\bf r}_i),
\end{equation}

where

\begin{equation}
{\bf a}_i=2\hbar\sum_{j\neq i}\frac{\hat z\times({\bf r}_i-{\bf r}_j)}
{|{\bf r}_i-{\bf r}_j|^2}
\end{equation}
is the Chern-Simons vector-potential. This Hamiltonian acts on the composite
fermion wavefunction $\Psi$ which is related to the electron wavefunction
$\Psi_e$
 through the phase factor:
\begin{equation}
\label{gauge}
\Psi_e=\prod_{i < j}\frac{(z_i-z_j)^2}{|z_i-z_j|^2}\Psi,
\end{equation}
where $z=x+iy$ is a complex coordinate.
 The self--consistent Hamiltonian for the gauge field problem has been
derived in the context of anyon superconductivity.\cite{Hanna,Halperin92}
Following the derivation of Halperin\cite{Halperin92} the  Hartree
Hamiltonian can be written as
\begin{equation}
H=\frac{({\bf p}+e/c{\bf A}({\bf r})-\langle {\bf a}({\bf r})
\rangle)^2}{2m^*} + U_j({\bf r}) + U_c({\bf r}),
\end{equation}
where $m^*$ is the effective mass. Although in Hartree approximation $m^*=m_B$
 \cite{HLR} I use a different value of $m^*$ that yields correct energy
 gaps for the
 fractional quantized Hall states (see bellow). This simple substitution
 should capture correctly short-range features in electron density related to
 the fractional  channels. A defficiency at large lengthscales can be fixed by
 an insignificant monotonic redistribution of the background charge.

The expectation value of the Chern-Simons vector-potential is given by
\begin{equation}
\langle{\bf a}({\bf r})\rangle =2\hbar\int d^2{\bf r}'\frac{\hat
 z\times({\bf r}-{\bf r}')}{|{\bf r}-{\bf r}'|^2}\langle \rho({\bf r}')
\rangle,
\end{equation}
with the particle density operator
\begin{equation}
\rho({\bf r}) = \sum_i \delta({\bf r}-{\bf r}_i).
\end{equation}
The Chern-Simons contribution to the scalar potential is
\begin{equation}
U_j({\bf r})=2\frac{\hbar}{e}\int d^2{\bf r}'\frac{\hat z\times({\bf r}-{\bf
r}')}
{|{\bf r}-{\bf r}'|^2}{\bf j}_H({\bf r}'),
\end{equation}
where the Hartree approximation to the expectation value of the current density
is
\begin{equation}
{\bf j}_H({\bf r})=\langle{\bf j}_p({\bf r})\rangle-\frac{e}{m^*}
\langle\rho({\bf r})\rangle (\langle{\bf a}({\bf r})\rangle-{\bf A}({\bf r})),
\end{equation}
while
\begin{equation}
{\bf j}_p({\bf r})=\frac{e}{2m^*}\{{\bf p}_i,\delta({\bf r}- {\bf r}_i)\}
\end{equation}
Finally the Coulomb interaction is included in
\begin{equation}
U_c({\bf r})=\frac{e^2}{\epsilon}\int d^2{\bf  r}'\frac{\langle\rho({\bf r}')
\rangle -\rho_+({\bf r}')}{|{\bf r}- {\bf r}'|},
\end{equation}
where $\rho_+({\bf r})$ is the positive background density.

In the chosen geometry the Hamiltonian can be simplified further taking
advantage
of the translational invariance in $y$ direction and
using dimensionless units written as
\begin{eqnarray}
\label{har}
H&=&-\frac{1}{2}\frac{d^2}{dx^2}
+\frac{1}{2}(k_m+x-\int dx' {\rm sgn}(x-x')\nu(x'))^2\nonumber\\
&+&\int dx'{\rm sgn}(x-x')\sum_{m,l} \nu_{m,l}(x')(k_m+x'\nonumber\\
&-&\int dx'{\rm sgn}(x'-x'')\nu(x''))\nonumber\\
  &+&\alpha\int dx' {\rm ln}(x-x')[\nu(x')-\nu_+(x')],
\end{eqnarray}
where $k_m$ is the momentum of the $m$-th state, $l$ is the Landau level index,
$\nu$ is the electron filling factor. The distances are in units of the
magnetic
length, $l_H$. The energy is measured in the units of the cyclotron frequency
 for composite fermions at $\nu=1$.
The strength of the Coulomb interaction compared to the kinetic energy
 is determined by
\begin{equation}
\alpha=\frac{1}{\pi}\frac{m^*}{m_B}\frac{l_H}{a_B}.
\end{equation}
Large $\alpha$ makes deviations of the electron density from the positive
background very
costly. This means that despite the high kinetic energy electron density mimics
the
positive background. For small values of $\alpha$ the kinetic energy of
composite fermions
becomes important leading to the formation of the fractional plateaus. For the
experimentally relevant parameters $m^*=4m_B$, $l_H=a_B=100$\AA $\;$ the value
of $\alpha\approx 1$
is found. I use $\alpha=1$ throughout this paper except for Sec.\
\ref{sec:res1/3}. Introduction of a short distance cut-off in the logarithm
in Eq.(\ref{har}) reflecting a finite $z$-extent of the wavefunctions was
found to have a similar effect as the reduction of $\alpha$.

This self-consistent Hamiltonian(\ref{har}) was solved numerically in this work
and
independently by Brey\cite{Brey}. Alternatively one may view the solution as a
variational
electron wavefunction of the form (\ref{gauge})
where $\Psi$ is a product of one-particle wavefunctions and all the
correlations are taken care of by the phase factor.

\section{Results for simple fractions $\nu=1/(2k+1)$.}
\label{sec:res1/3}
The special role of the filling fractions  $\nu=1/(2k+1)$ in the FQHE has been
recognized
ever since Laughlin\cite{Laughlin} proposed his wavefunctions. The knowledge of
the explicit wavefunctions made possible the calculation of the density
profile for a  $\nu=1/(2k+1)$ state.\cite{Mitra,Rezayi} In this section I
compare
results of the Hartree calculation with the calculations for the Laughlin
wavefunction.\cite{Rezayi}

As shown by Wen\cite{Wenb} a  $\nu=1/(2k+1)$ state  in a very sharp confinement
supports a single branch of edge excitations. In terms
of composite fermions this implies that there is a single energy band, giving
rise to a single edge channel. Since the composite fermion energy spectrum
does not have an intuitive meaning in the electron representation I can only
compare the density distribution.

\begin{figure}
\epsfxsize=5truein
\hskip 0.0truein \epsffile{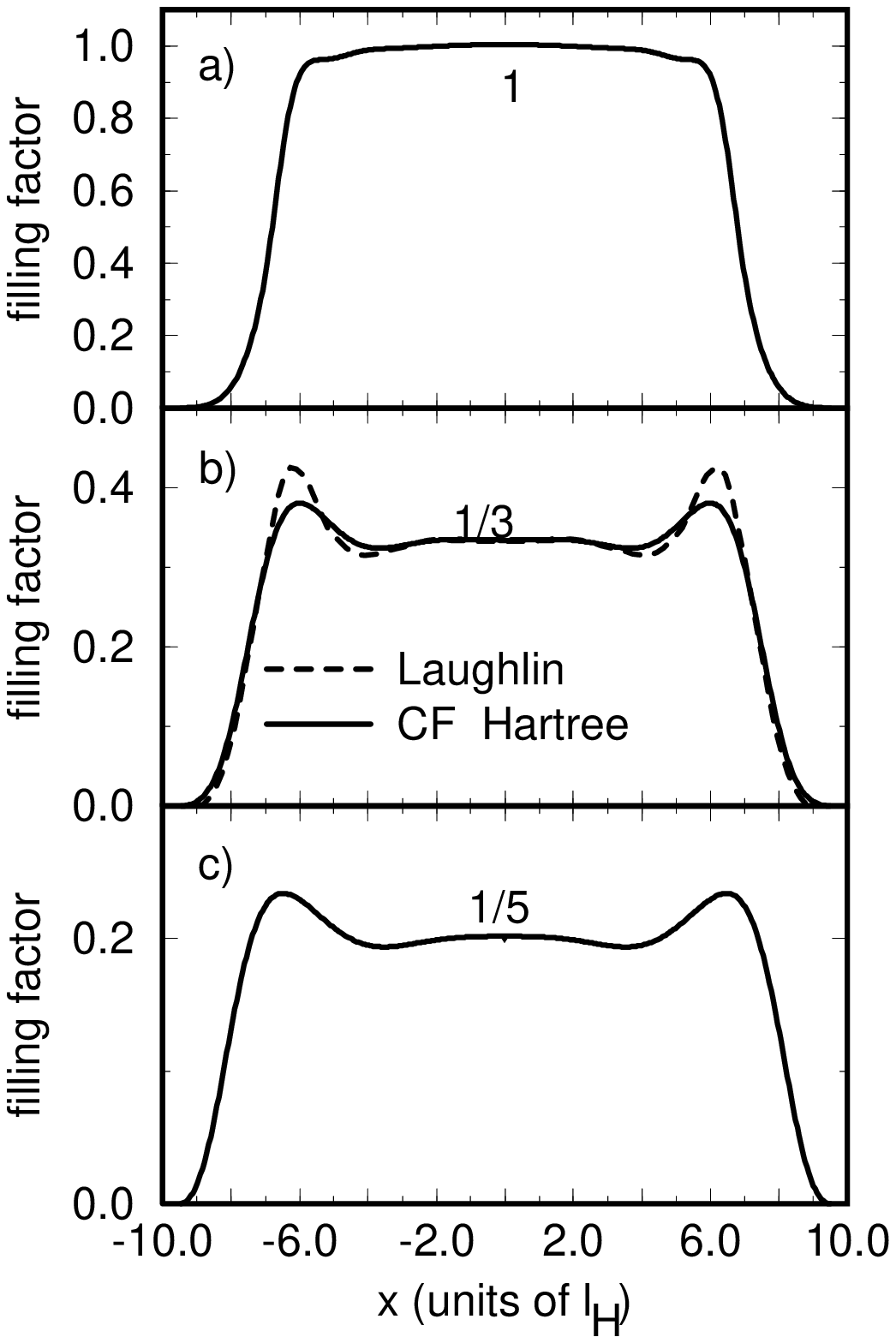}

{ FIG 3: Electron density distribution obtained from the composite
 fermion Hartree calculation for simple
 filling fractions  in a very sharp confinement.
a) Filling factor $\nu=1$ state.
b) Filling factor $\nu=1/3$ state. Dashed line: Rezayi-Haldane calculation for
Laughlin wavefunction;
Full-line: composite fermion Hartree calculation.
c) Filling factor $\nu=1/5$ state.}
\label{fig:simple}
\end{figure}

In Fig.\ref{fig:simple} I present the electron density
distribution for three simple fractions in the case of the sharp edge.
In the case of $\nu=1$ the density is almost featureless and is very close to
the
 profile expected for the lowest Landau level filled up to the Fermi momentum.
The density distribution at fractional filling factors shows damped
oscillations
near the edge.  These oscillations have been first found
numerically\cite{Zhang,Rezayi85}.
Their period corresponds to the inter-particle spacing. The oscillations were
argued to be the precursor of the Wigner solid formation and are reproduced in
the
 single-mode approximation to the density-density response
function.\cite{Girvin}

It is clear that the composite fermion Hartree calculation is able to reproduce
the
 essential features in the density distribution such as the period and the
damping
of the oscillations. In the following sections the method is used to study the
edge
structure when the exact wavefunction is not available.

The composite fermion approach allows oscillations in density to be related to
a pole in
the Fourier transform of
the density-density response function (static susceptibility) $K_{00}(q)$. To
demonstrate
this I consider linear response to a potential of the form
\begin{equation}
V(x,y)=V_0 \delta(x).
\end{equation}
The induced change in the electron density is then independent of $y$ and
expressed as
\begin{equation}
\delta\rho(x)=V_0\int_{-\infty}^{+\infty}dq K_{00}(q)e^{iqx}
\end{equation}
The large $x$ behavior of $\delta\rho(x)$ is dominated by a pole in the complex
$q$-plane closest to
the real axis. For a pole at $q=q'+iq''$ the asymptotic form of the induced
change in the
electron density is
\begin{equation}
\label{oscil}
\delta\rho(x)\sim e^{-q''x}cos(q'x)
\end{equation}

The electromagnetic response function for the fractional states has been
calculated by Simon and
Halperin\cite{Simon} by using the RPA for composite fermions. I use their
results neglecting
the effect of Coulomb interaction and the renormalization of the composite
fermion mass.
 When there is only one filled band for composite fermions the static
density-density response
 function is
\begin{equation}
\label{K_00}
K_{00}(q)=\frac{-q^2\Sigma_0}{2\pi \hbar \Delta \omega_c
 \bigl(\bigl(\tilde\phi
\Sigma_1+1\bigr)^2-\tilde\phi^2\Sigma_0\bigl(\Sigma_2+1\bigr)\bigr)}
\end{equation}
where the number of attached flux quanta $\tilde\phi=-2k$ for the fractions
$1/(2k+1)$,
 $\Delta \omega_c=\omega_c/(2k+1)$ is the cyclotron frequency for composite
fermions and
$\Sigma_j$ are functions of $q^2$ as defined in Ref.\cite{Simon}.
By performing an analytic continuation of Eq.\ref{K_00} to the complex plane
one can see that
this function has a pole at $q\approx (1.3+0.6i)l_H^{-1}$ for $\tilde\phi=-2$
corresponding
to $\nu=1/3$ and $q\approx (1.1+0.5i)l_H^{-1}$ for $\tilde\phi=-4$
corresponding
to $\nu=1/5$. This pole leads to damped
oscillations in the induced change in the electron density of the form
(\ref{oscil}).
The inclusion of  Coulomb interaction and the renormalization of the composite
fermion mass
shifts the position of the pole in the complex plane but the qualitative
picture remains the same.

\section{Results for the $\nu=1$ edge.}
\label{sec:res1}

\begin{figure}\epsfxsize=5.5truein
\hskip 0.0truein \epsffile{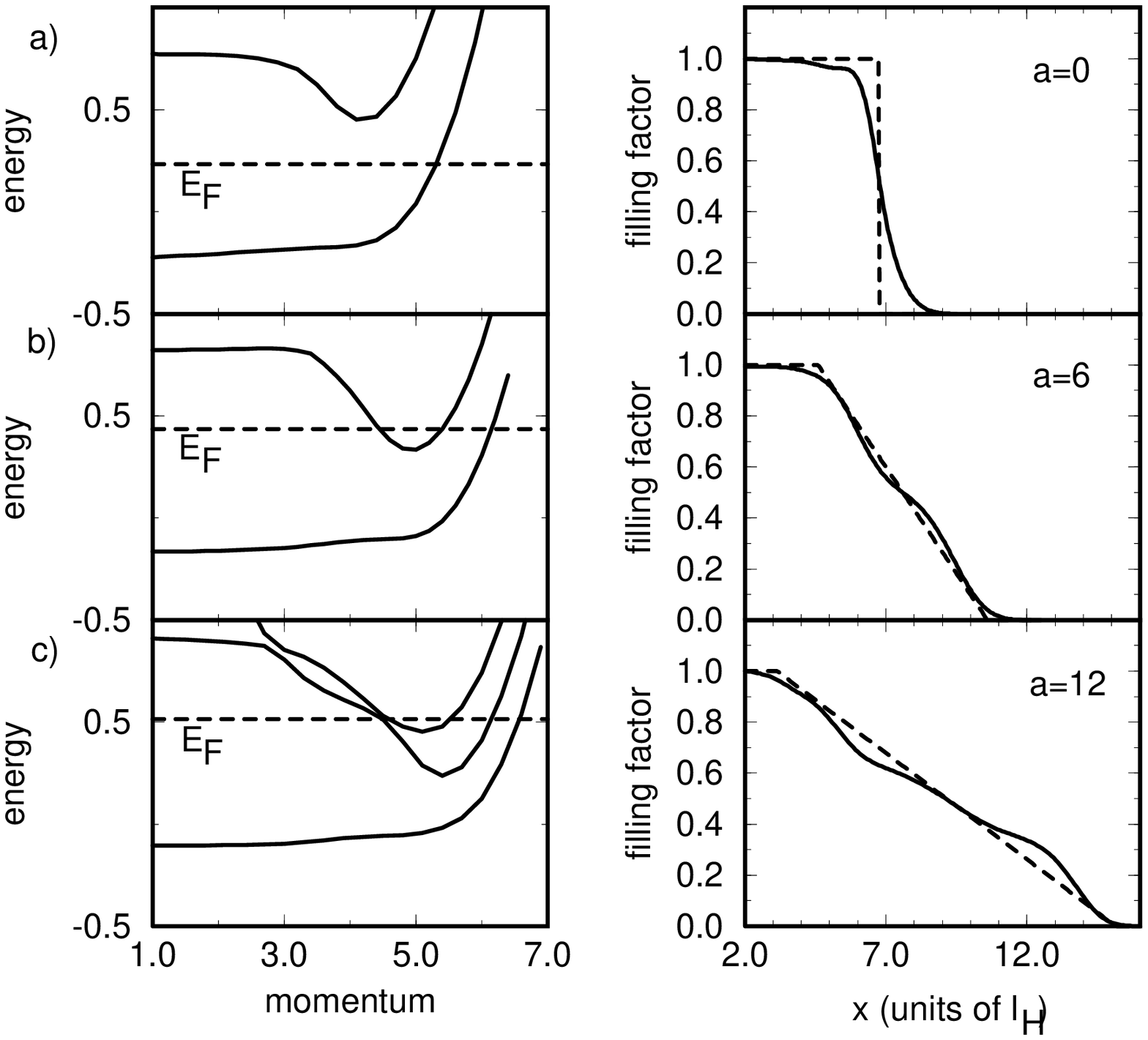}

 { FIG 4: Results of the Hartree calculation for the $\nu=1$ edge for different
sharpness of the confinement
potential parameterized by $a$. Composite fermion energy spectrum is in the
left column (dashed
line denotes the Fermi level), and the electron density distribution is in the
right column
(dashed line shows the neutralizing background).
a) $a=0$. b) $a=6l_H$. c) $a=12l_H$}
\label{fig:nu1}
\end{figure}

The properties of the electron state at the Landau level filling factor $\nu=1$
in the bulk
 can be understood mostly from the standpoint of non-interacting electrons.
This may not
be the case at the edge of the system where the filling factor falls slowly to
zero as demonstrated
in experiments with non-ideal contacts.\cite{Kouwenhoven2/3} The detailed
structure of the
edge is determined by the strength of the confinement potential since it
dictates
the filling factor gradient at the edge. The approach outlined in
Sec.\ref{sec:model} enables
 me to study the structure of the $\nu=1$ edge  in terms of non-interacting
composite fermions.
 Here I give a physical interpretation to the results of the Hartree
calculation shown in
 Fig.\ref{fig:nu1}. (Plots are restricted to the right edge only because of the
symmetry of the
problem.)

In the case of a very sharp edge($a\approx l_H$) the confinement energy
dominates over the
 electron-electron interactions. Consequently, the one-electron picture gives
an adequate
description of the edge: a single one-dimensional edge channel is formed at the
intersection
of the Landau level and the Fermi level. In the composite fermion description
the $\nu=1$
state corresponds to a single filled Landau level in the reversed effective
magnetic field.
 At the edge this level intersects the
 Fermi level forming a single channel as illustrated in Fig.\ref{fig:nu1}a. The
higher
composite fermion Landau levels remain unfilled.

 The self-consistent energy spectrum obtained in the Hartree calculation can be
understood
 by solving a one-particle problem.\cite{Chklovskii3,Lee} Since the effective
magnetic
field follows the electron density profile, composite fermions find themselves
in a step-like
magnetic field.  In the Landau gauge the vector potential is $A=B_0|x|$, where
$x$ is the
 distance from the edge. The problem is reduced to a one-dimensional
Schr\"odinger equation
similar to the uniform magnetic field case. However the effective potential in
this
equation has two potential wells at large momenta. This leads to the existence
of a degenerate
pair of states centered at positive and negative $x$. The degeneracy is lifted
by the
confinement potential. But it is important to remember
that the momentum of a given state only determines its distance from the zero
magnetic field
line. States with the same momentum may be on different sides of this line.
To get the idea of the energy spectrum in the real space it is useful to
reflect
every other energy band around the intersection of the lowest one with the
Fermi level.

For larger $a$ there must be a finite width region with the filling factor
between
one and zero. In the one-electron approximation this implies a huge degeneracy
of the
states at the edge. The composite fermion approach allows me to resolve this
problem. The self-consistent Hartree energy spectrum for $a=6l_H$ is presented
in
Fig.\ref{fig:nu1}b.
The main difference from the sharp confinement case is the descent of the
second
Landau level below the Fermi energy. Again this result follows from the
solution of
the one-particle problem for composite fermions. The effective magnetic field
variation
may be approximated by a constant gradient. The spectrum obtained by solving
this
problem\cite{Muller,Chklovskii3,Lee} with the confinement potential is very
close to the
one in Fig.\ref{fig:nu1}b. The descent of the Landau level is because of the
reduction
of the cyclotron energy in the vicinity of $\nu=1/2$. As the composite
fermion band touches the Fermi level a pair of
\twocolumn[\hsize\textwidth\columnwidth\hsize\csname @twocolumnfalse\endcsname
] fractional channels is born.
 It is
not clear how to identify these channels with any particular filling fraction
since there
is no incompressible state in between.

For larger $a$ higher composite fermion bands become occupied giving rise to
additional pairs of channels. Incompressible states are formed between the
channels as
shown in Fig\ref{fig:nu1}c. When the incompressible regions are wide the energy
spectrum
in the momentum space can be mapped to the real space. Since the effective
magnetic
 field for composite fermions changes sign at $\nu=1/2$ the dependence of
coordinate on momentum is double-valued. The position of the states belonging
to
the odd-numbered bands is given by their momentum up to the factor of magnetic
length.
The position of the states in the even-numbered bands can be obtained roughly
by reflecting
these bands around the $\nu=1/2$ line (see Fig.\ref{fig:schematic}). This
allows me to
 identify the channels by the filling fractions between them. The second band
gives rise
 to the $1/3$ channel (between $\nu=0$ and $\nu=1/3$) while the third band
produces the
$2/3$ channel (between $\nu=2/3$ and $\nu=1$). According to the composite
fermion picture
there are also three channels between $\nu=2/3$ and $\nu=1/3$.

As shown in Fig.\ref{fig:nu1} the composite fermion approach fractional
channels are born
 in pairs as the confinement is relaxed in agreement with the prediction by
Wen\cite{Wen}.
The formation of the first pair of the fractional edge  channels is reminiscent
of the
edge reconstruction proposed by Chamon and Wen\cite{Chamon}. However, the
physics is
 very different here. Electron-electron {\it correlations} are responsible for
the
 formation of the {\it fractional} channels as opposed to the {\it exchange}
interaction
 producing a pair of {\it integer} channels in the edge reconstruction
model\cite{Chamon}.
In order to determine the relationship between the two effects a solution
that takes into account both exchange and correlations is needed.

As the edge becomes very smooth the one-particle composite fermion approach
must break down
because the filling factor is expected to vary smoothly between any two
principal
sequence filling factors. This calls for the transformation to the next
generation of composite fermions. A simple calculation shows that a range
of confinement strength exists where the one-electron description fails
 while the one-particle approach for composite
fermions of the principal sequence is still valid. The main reason is
that the composite fermions of the higher generation see a stronger gradient of
the effective magnetic field at the same density gradient. The calculation
 is similar to the one given for higher Landau levels in
Sec.\ref{sec:classical}.
\twocolumn[\hsize\textwidth\columnwidth\hsize\csname @twocolumnfalse\endcsname
\begin{figure}
\epsfxsize=5.5truein
\hskip 0.0truein \epsffile{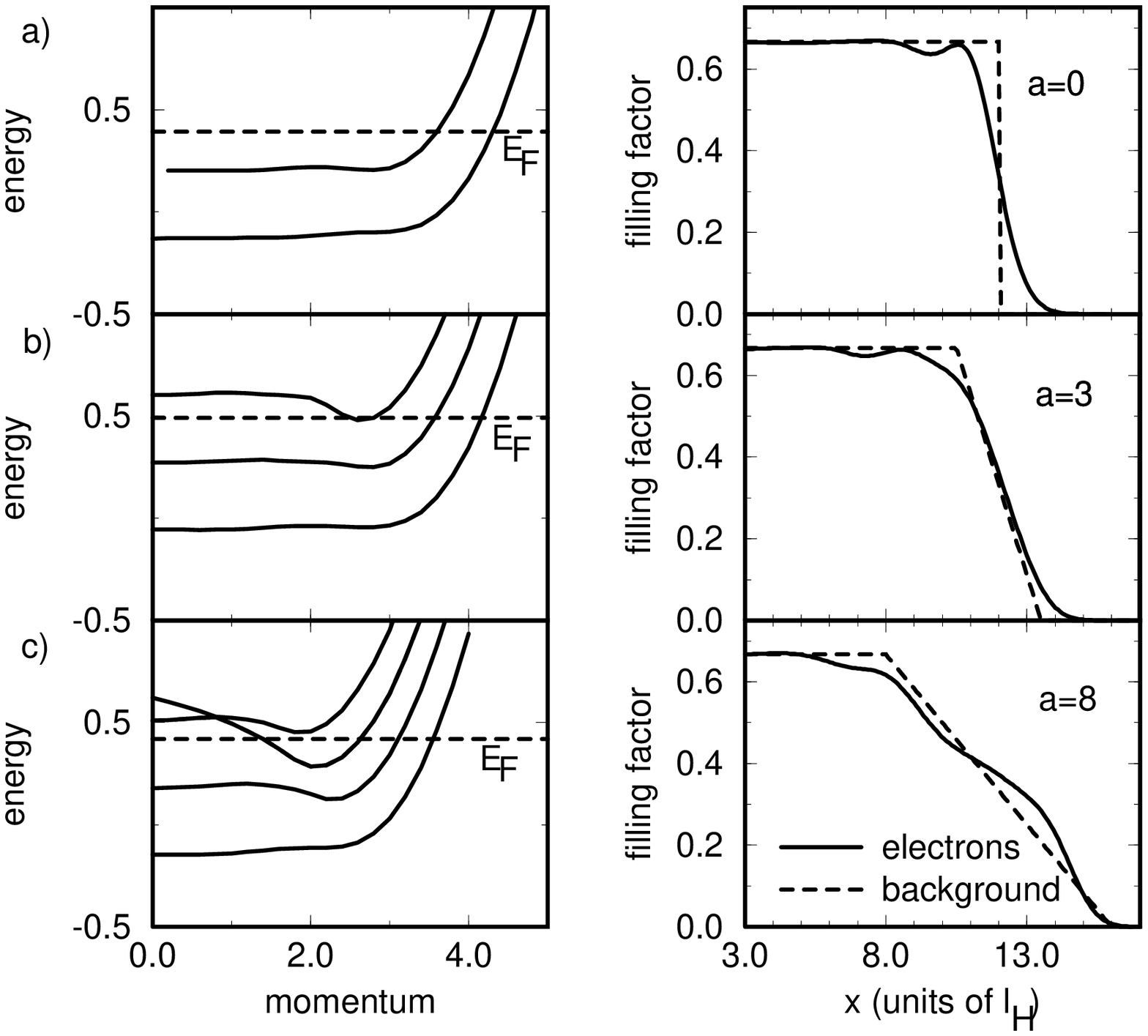}

{ FIG 5: Results of the Hartree calculations for the $\nu=2/3$ edge
  for different
sharpness of the confinement
potential parameterized by $a$. Note the transition from the MacDonald to the
Chang-Beenakker picture
as the confinement is relaxed.
a) $a=0$. b) $a=3l_H$. c) $a=8l_H$.}
\label{fig:res2/3}
\end{figure}]

At this point I would like to address the effect of disorder on the edge
 structure. In the case of short-range disorder fermions may scatter
between different channels making it a complicated problem. However, when
disorder is long-ranged a simple model emerges. A superposition of
the slowly varying disorder potential and the confinement potential
can be modeled by a background charge with width $a$ varying slowly along the
edge. By treating the problem adiabatically one gets the number of channels
 varying along the edge. It is clear that the channels corresponding to
the higher effective Landau levels become localized in the regions of
smooth edge (large $a$). The number of delocalized channels is determined
by the steepest confinement region (smallest $a$).

\section{Results for the $\nu=2/3$ edge.}
\label{sec:res2/3}
 Wen has shown\cite{Wenb} that the $\nu=2/3$ quantum Hall state must
support more than one branch of edge excitations. Two, seemingly incompatible,
 theoretical models have been proposed to explain the detailed structure
 of the composite edge. The first model due to MacDonald\cite{MacDonald} claims
 the existence of two excitation branches and the absence of the  $\nu=1/3$
state at the edge. The second model was proposed by  Beenakker\cite{Beenakker}
 and Chang\cite{Chang}. In their picture there is a transition
from the 2/3 state to the 1/3 state and from the 1/3 state to 0 with
compressible
regions in between.

I use the composite fermion approach to address this controversy. Results of
the
calculation are shown
in Fig.\ref{fig:res2/3} for different sharpness of the confinement potential.
In the case of the sharp edge, $a=0$, there are clearly two edge channels in
agreement
with MacDonald's model. However, I do
not find any region with the filling factor close to one. Therefore it is not
clear how
to identify these channels.

As the confinement is relaxed, Fig.\ref{fig:res2/3}b, the third composite
fermion
energy band descends and crosses the Fermi level. This signals the formation of
a
 pair of edge channels. Further smoothing of the confinement leads to the
formation of the incompressible  $\nu=1/3$ state, Fig.\ref{fig:res2/3}c.
 The electron density distribution is in
agreement with the Chang--Beenakker picture. One  can see incompressible
regions
corresponding to $\nu=1/3$ and $\nu=2/3$ and compressible regions between them.
However, according to the composite fermion approach there are four channels in
this case:
one of them is located between $\nu=1/3$ and $\nu=0$, and other three are
between  $\nu=1/3$
and $\nu=2/3$.

As the edge becomes smoother higher bands must  get occupied reflecting the
formation
 of new fractions at the edge in analogy with the
$\nu=1$ edge discussed in Sec.\ref{sec:res1},\ref{sec:classical}.

\section{Quasiclassical approximation.}
\label{sec:classical}
For a very smooth edge, $a\gg l_H$, the number of filled composite
fermion bands is large, allowing me to use the quasiclassical
approximation to find the number of fractional channels. To be specific
I consider the edge of the $\nu=1$ state and restrict myself to
the fractions of the principal sequence $p/(2p+1)$.\cite{Jain}
The question may be restated as: what is the value of $p$ for the
highest fractional Hall state
that survives the density gradient determined by $a$?
The value of $p$ is given by the highest filled energy band of
composite fermions. To answer this question I look for the classical
orbit of composite fermions moving with the Fermi velocity and
enclosing the largest number of the effective magnetic field flux
quanta. The number of enclosed flux quanta gives the value of $p$.

This argument is similar to the estimate of the last integer Hall plateau in a
quantum wire by Roukes {\it et al.}.\cite{Roukes} They have argued that the
maximum
number of edge channels is given by the number of flux quanta
encircled by the largest cyclotron orbit that fits in the wire.
\begin{figure}\epsfxsize=3truein
\hskip 0.0truein \epsffile{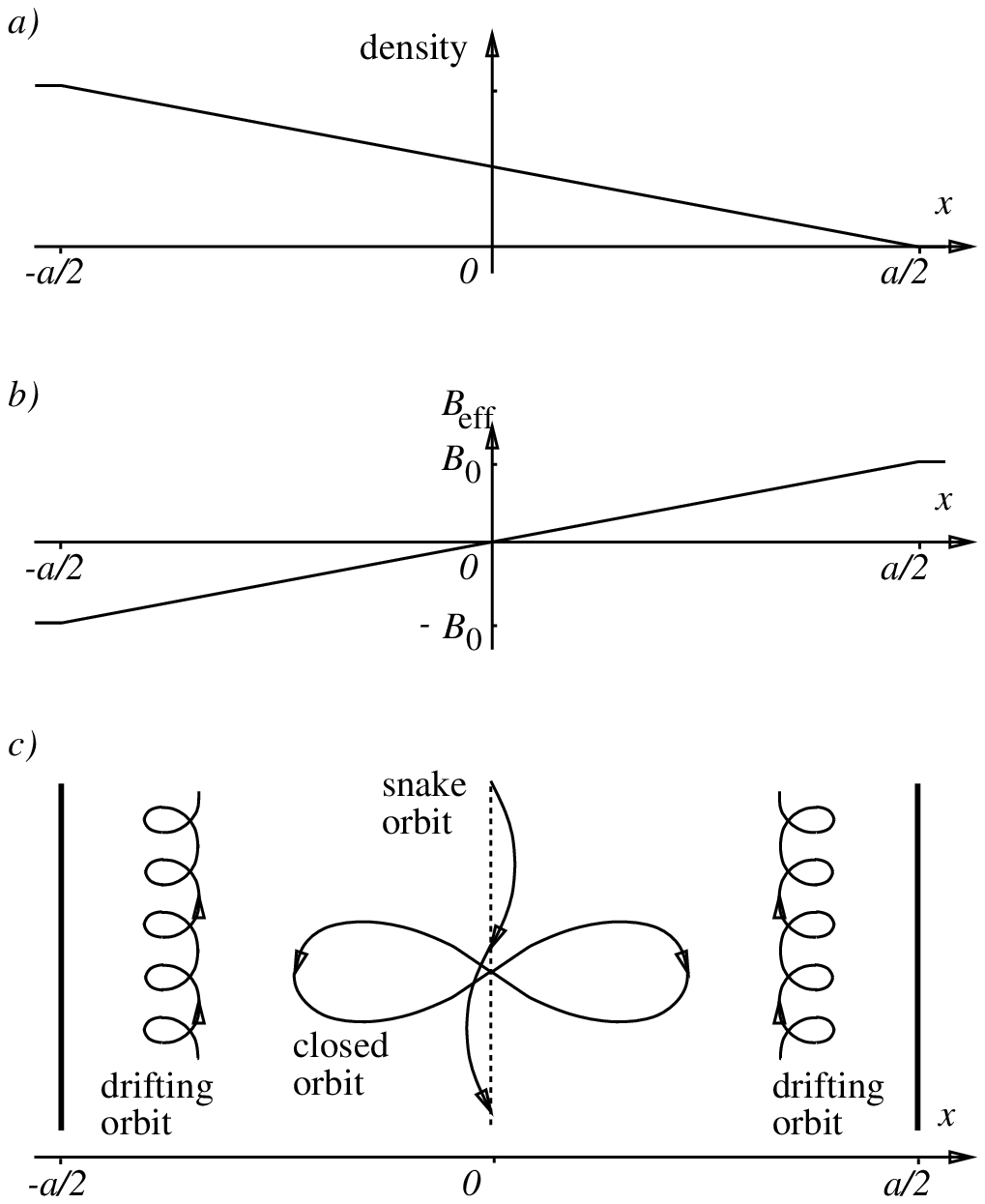}

{ FIG 6: Quasiclassical approximation (dependence of the Fermi velocity on
the density is neglected).
 a) Electron density profile in the approximation of the constant gradient.
b) The effective magnetic field seen by composite fermions.
c) Three kinds of classical orbits possible in the varying effective magnetic
field.}
\label{fig:classical}
\end{figure}
Assuming that the electron density roughly follows the positive background
(Fig.\ref{fig:classical}a), composite fermions see a roughly linearly varying
 effective magnetic field, Fig.\ref{fig:classical}b:
\begin{equation}
B(x)\approx B_0 x/a,
\end{equation}
where $x$ is the distance from the zero effective field ($\nu=1/2$) line.
 There are three types of classical orbits for composite fermions in this
field\cite{Muller,Chklovskii3,Lee}: drifting orbits, that move along the  edge
 in the direction of electron drift, snake orbits that move in the opposite
direction
and closed orbits that do not drift along the edge, Fig.\ \ref{fig:classical}c.
I focus on the closed orbits which are the classical analogue of the states
 at the minima of the energy bands and must be filled first.

Neglecting the dependence of the Fermi velocity on the deviation of the filling
factor from $\nu=1/2$ the composite fermion cyclotron radius as a function of
$x$:
\begin{equation}
\label{cyc}
R_c(x)\sim l_H a/|x|
\end{equation}
where I used the fact that the cyclotron radius at the boundary of the
compressible
strip ($x=-a$) is $l_H$. The closed orbit should have the extent in the $x$
direction
 such that $R_c$ at $x$ is equal  to $x$. By using Eq.(\ref{cyc})
 I find from this condition that
\begin{equation}
\label{cyclotron}
R_c\sim \sqrt{l_H a}.
\end{equation}
The number of channels is equal to the number of the effective
flux quanta enclosed by the orbit. The area of the orbit is proportional to
$R_c^2$.
The typical area corresponding to one flux quantum at $|x|=R_c$ is $l_H^2
a/R_c$.
 Then the number of flux quanta enclosed by the orbit is
\begin{equation}
\label{square}
p\sim \frac{R_c^2}{l_H^2a/R_c}
\end{equation}
By using the value of $R_c$ from Eq.\ref{cyclotron} I  find the number of
channels for the smooth edge of the width $a$:
\begin{equation}
\label{result}
p\sim \sqrt{\frac{a}{l_H}}
\end{equation}
This result agrees with the work of Lee, Chalker, and Ko\cite{Lee}
who used the WKB approximation to study the motion of a particle in a slowly
varying magnetic field.

Eq.(\ref{result}) counts only the channels for composite fermions of the
principal sequence.
In the case of a very smooth edge composite fermion channels must become wide.
Then one should make a transformation
to the composite fermions of the next generation and follow the same line of
argument.

If one views transport in the QHE as percolation along
the network of edge states in the bulk, Eq.(\ref{result}) puts an upper
bound on the highest fractional Hall state that can be observed.
In this case $a$ is determined by the typical density gradient along
the percolation line in the bulk.

The above argument can also be generalized to higher Landau levels.
 Then $a$ denotes the
distance where the filling factor changes by one. Therefore if the magnetic
 field is varied while density is fixed, the number of channels in each
Landau level is
\begin{equation}
p_{N}=p_{1}/N^{3/4},
\end{equation}
where $p_{1}$ is given by Eq.(\ref{result}) and $N$ is the bulk filling factor.
This result may account for difficulties in observing fractional Hall
states in higher Landau levels. When the filling factor is close to $N-1/2$
the highest fractional Hall state is given by
\begin{equation}
p_{N-1/2}=p_{1/2}/(2N-1)^{3/4},
\end{equation}
where $p_{1/2}$ gives the highest fraction in the lowest Landau level.

\section{Summary.}

In this paper I have presented a theory of the compressible liquid strip at the
edge of an incompressible quantum Hall state. The theory is based on the
 composite fermion approach and describes the evolution of the edge as
the steepness of the confinement potential is varied.
When the width of the compressible strip is of the order
of several magnetic lengths the composite fermion Hartree approximation
 is used to find numerically the density distribution and the energy
 spectrum for composite fermions.
 This allows me to predict the number of fractional channels for a given
confinement potential.
 The direction of propagation and the detailed structure of the excitations in
these channels require further study.

 In the asymptotic limit of a wide compressible strip the number of channels
can be found quasiclassically. It is given by Eq.\ref{result}, assuming
that the density gradient is roughly constant. The result gives an upper
bound on the highest fraction that can be observed in the FQHE for a
given strength of long-range disorder.
 Damped density oscillations at the edge are found for filling factors
$\nu=1/(2k+1)$ in the case of a sharp confinement. They are related to the pole
in the  linear response static susceptibility derived in RPA for composite
fermions.
The $\nu=2/3$ edge structure is found to depend strongly on the steepness of
the confinement potential. The transition from the MacDonald to the
Chang-Beenakker picture is observed when the width of the edge is $\sim 3l_H$.

\acknowledgments

I am grateful to B. I. Halperin for numerous helpful suggestions and careful
reading of the
manuscript, and to F. G. Pikus for advice on computer simulation.
 I would like to thank L. Brey as well as the authors of
Refs.\cite{Alphenaarpre,Lee,Rezayi} for
sending their preprints prior to publication and N. Cooper for comments and
corrections in the
manuscript. I have benefited from discussions with
 P. A. Lee, X.-G. Wen, B. Y. Gelfand, D. K. K. Lee, K. A. Matveev, B. I.
Shklovskii, S. H. Simon, and A. Stern.
This work has been supported by NSF Grant No. DMR92-16007 at MIT and by a
Junior Fellowship
of the Society of Fellows and by NSF Grant No. DMR91-15491 at Harvard
University.
 I acknowledge the hospitality of the AT\&T Bell Laboratories.

\end{document}